# On the limits to mobility in InAs quantum wells with nearly lattice-matched barriers

B. Shojaei[1], A.C.C. Drachmann[2], M. Pendharkar[3], D.J. Pennachio[1], M.P. Echlin[1], P.G. Callahan[1], S. Kraemer[1], T.M. Pollock[1], C.M. Marcus[2], and C.J. Palmstrøm[1, 3, a]

1.  Materials Department, University of California, Santa Barbara, CA 93106, USA

2.  Center for Quantum Devices, Niels Bohr Institute, University of Copenhagen, 2100 Copenhagen, Denmark

3.  Department of Electrical and Computer Engineering, University of California, Santa Barbara, CA 93106, USA

*Abstract*

The growth and the density dependence of the low temperature mobility of a series of two-dimensional electron systems confined to un-intentionally doped, low extended defect density InAs quantum wells with $Al_{1-x}Ga_xSb$ barriers are reported. The electron mobility limiting scattering mechanisms were determined by utilizing dual-gated devices to study the dependence of mobility on carrier density and electric field independently. Analysis of the possible scattering mechanisms indicate the mobility was limited primarily by rough interfaces in narrow quantum wells and a combination of alloy disorder and interface roughness in wide wells at high carrier density within the first occupied electronic sub-band. At low carrier density the functional dependence of the mobility on carrier density provided evidence of coulombic scattering from charged defects. A gate-tuned electron mobility exceeding 750,000 $cm^2$/Vs was achieved at a sample temperature of 2 K.

__________________________

a)  Electronic mail: cpalmstrom@ece.ucsb.edu

*Manuscript*

A significant achievement of material science is the perpetual enhancement of low temperature two-dimensional (2D) electron and hole mobility in semiconductor heterostructures. Modulation doping, a technique that allows the spatial separation between dopants and carriers[1], and repeated advances in molecular beam epitaxy (MBE) have led to carrier mobility greater than $10^7$ cm$^2$/Vs and greater than $10^6$ cm$^2$/Vs in electron[2,3] and hole[4] doped GaAs based 2D systems, respectively. These achievements are limited to GaAs-based heterostructures; similar advances have yet to be made in other compound semiconductor heterostructures.

Of particular interest are heterostructures wherein carriers are confined to narrow gap semiconductors such as InAs and InSb. Strong spin-orbit coupling in narrow gap semiconductors and their heterostructures make them suitable for developing novel spin based electronics[5–7] and for realizing topological superconductivity in semiconductor-superconductor heterostructures[8–11]. For their utilization in quantum information processing low-dimensional systems require sufficiently high carrier mobility[12]. In state-of-the-art GaAs based 2D electron systems, it is believed mobility is limited by homogeneous background impurities[13]. If similar scattering rates could be achieved in the narrow gap semiconductors, then a scaling by the effective mass would yield higher mobility than the best GaAs 2D electron systems, by a factor of approximately 3 for carriers confined to InAs and approximately 4 for carriers confined to InSb. Reported values of mobility[14,15] are a factor of 100 lower than what may be theoretically possible for the narrow gap semiconductors. Advances require a greater understanding of the scattering mechanisms in heterostructures of the narrow gap semiconductors.

This work reports on scattering mechanisms and the limits to mobility in multiple InAs quantum well heterostructures with varying well thicknesses and interface growth procedures in which the density dependence of the low temperature mobility was determined by magnetotransport experiments using a dual gated Hall bar device geometry. In 15 nm wide InAs quantum wells a non-monotonic dependence of mobility on electron density was observed over the range $2 \times 10^{11}$ cm$^{-2}$ to $1.8 \times 10^{12}$ cm$^{-2}$. A gate-tuned electron mobility exceeding 750,000 cm$^2$/Vs was achieved at a sample temperature of 2 Kelvin. The possible scattering mechanisms leading to the observed density dependence of the mobility are discussed.

Heterostructures were grown by molecular beam epitaxy (MBE) on GaSb:Te (001) substrates. Growth rate and composition calibrations were performed prior to the growth of the heterostructures and verified by reflection high energy electron diffraction patterns and oscillations during the growth of the heterostructures. Fig. 1(a) depicts a schematic MBE grown structure with integrated gate dielectric and front gate metallization. The conduction band in the vicinity of a 15 nm wide InAs quantum well is depicted in Fig. 1(b) for a well with $Al_{0.8}Ga_{0.2}Sb$ barriers. The self-consistent calculation depicts a case wherein the Fermi energy lies between the first and second electronic sub-band of the two dimensional system. For the structures used in this study, the Fermi energy could be gate tuned using the dual gated Hall bar geometry shown in Fig. 1(c), where the tellurium doped GaSb substrate serves as one electrostatic gate and the Ti/Au metallization over the $Al_2O_3$ dielectric serves as a second electrostatic gate.

Five structures, denoted samples A through E, were grown using similar electrically isolating $Al_{1-x}Ga_xAs_{1-y}Sb_y$/AlSb buffer layers through the AlSb/GaSb superlattice using a cumulative group V:III beam equivalent pressure (BEP) ratio of 4:1 and a substrate temperature of 510 °C. The substrate temperature was reduced to 470 °C for the growth of the superlattice and the remaining layers. The barriers and InAs layer were grown using a BEP ratio of 3:1 and 4:1, respectively. The set of structures comprised variations in the InAs layer thickness, the shutter sequence used to transition the growth between the barriers and quantum well, and the composition of the $Al_{1-x}Ga_xSb$ barriers. Growth of samples A, B and C made use of a shutter sequence, denoted Procedure I, developed by Tuttle and co-workers[16]. Growth of samples D and E implemented a shutter sequence, denoted Procedure II, wherein group V and III shutters were closed and opened simultaneously when the growth transitioned between barriers and the InAs layer. Table 1 summarizes several critical characteristics of samples A through E. Additional details of the growth and device fabrication can be found in reference 17[17].

The surfaces of the MBE-grown films appeared flat when examined under a Nomarski microscope at 100x magnification. Atomic force microscopy (AFM) of the surfaces revealed height variations corresponding to GaSb monolayer-height steps (0.3 nm) on all samples. Exemplar 20 x 20 $\mu m^2$ and 5 x 5 $\mu m^2$ AFM scans of sample E are shown in Fig. 2(a) and 2(b), respectively. Height variations along the [1-10] crystallographic direction starting at the triangular marker in Fig. 2(b) are depicted in Fig. 2(c).

In a previous work it was reported the carrier mobility and single-particle lifetime in InAs/(Al,Ga)Sb heterostructures are sensitive to the dislocation density in the epitaxial layers[18]. In this work the use of the nearly lattice matched substrate and buffer layers was intended to inhibit the formation of extended defects during MBE growth. Fig. 3(a) shows the thin film x-ray diffraction reciprocal space map in the vicinity of the 002 and asymmetric 115 Bragg reflections of the GaSb substrate of sample D. In the 115 reciprocal space map, the substrate peak and the peak from the $AlAs_{1-y}Sb_y$ buffer layer are well resolved and indicate the buffer was nearly lattice matched but slightly tensile strained to the substrate (y ~ 0.9). The threading dislocation densities of the films were measured by both etch pits using a solution of $HF:H_2O_2:H_2SO_4:H_2O$ similar to that reported by Aifer and Maximenko[19] and by electron channeling contrast imaging (ECCI) using a backscattering geometry in a field emission scanning electron microscope. ECCI analysis was performed using several diffraction vectors to mitigate unintentionally satisfying the invisibility criterion where the Burger's vector for a given dislocation is perpendicular to the diffraction vector. Fig. 3(b) shows an ECCI micrograph measured from the surface of the MBE grown film of sample D. Several lines of enhanced contrast are interpreted to be due to the presence of misfit dislocations confined to heterointerfaces within the epitaxial layers. ECCI analysis was also performed on a >1 μm GaSb buffer structure grown on a GaAs (001) lattice mismatched substrate. An ECCI micrograph of the buffer grown on the mismatched substrate, shown in Fig. 3(c), indicates a high density of surface penetrating threading dislocations which manifest as bright spots. For the quantum well heterostructures grown using lattice matched buffers both the ECCI analysis and etch pit densities indicate low threading dislocation densities; a conservative estimate of $10^6$ dislocations/cm$^2$ is used as an upper limit for all samples.

A high angle annular dark field (HAADF) scanning transmission electron microscope (STEM) image of sample D along the [110] zone axis and in the vicinity of the InAs quantum well is shown in Fig. 4. The approximately 80 nm thick specimen was prepared by milling a section of wafer using a focused gallium ion beam. The undulations at the interface between the superlattice and the AlAs$_{1-y}$Sb$_y$ buffer are interpreted to be due to strain in the buffer and a resulting change in its surface morphology during growth. The well and barrier layer thicknesses measured by HAADF-STEM in sample D are in agreement with the values expected from MBE growth rate calibrations.

Magnetotransport experiments were carried out on dual gated Hall bars at a sample temperature of 2 Kelvin in a He-4 cryostat and using an excitation current of 1 μA. Measurements of the Hall and longitudinal magnetoresistance, as shown for sample D in Fig. 5 at a front and back gate voltage of 0 V, allowed the determination of carrier density and carrier mobility. In all devices, the carrier density varied linearly with applied gate voltages. A gating efficiency of the front gate, $\frac{dn_s}{dV_f} \sim 3.5 - 4.0$ ($10^{11}$ cm$^{-2}$V$^{-1}$) was typical. The determination of carrier density by the slope of the low-field Hall resistance and by the periodicity of Shubnikov de Haas (SdH) oscillations were in good agreement, and are summarized in Table 1 at zero gate bias.

The carrier density dependence of the mobility was measured by gate-tuning the carrier density. To determine the scattering mechanisms that led to the functional dependence of the mobility on density a series of elastic scattering calculations were performed. These calculations included scattering due to background impurities, remote impurities, charged dislocations, interface roughness and alloy scattering[20–23]. The calculations are outlined in Appendix A and follow the treatment of the transport relaxation time outlined by Stern and Howard[24], where zero temperature is assumed, and inter-subband scattering, multiple scattering events, and correlations between ionized impurities are neglected. The calculations were intended to determine the expected functional dependence of the mobility on carrier density and qualitatively determine which scattering mechanisms may be dominant in InAs quantum wells with nearly lattice matched barriers over a given carrier density range. Transport relaxation times were calculated individually and the total mobility was evaluated using Mathiessen's rule.

The dependence of the mobility on density for samples A, B, and C is shown in Fig. 6(a). In all three cases, the density was tuned using the front gate. The back gate was fixed at 0 V and 0.8 V for sample A, and 0 V for samples B and C. For the 10 nm wide well, sample B, and the 8 nm wide well, sample C, the mobility increased monotonically with density. The dependence of mobility on density in the 15 nm wide well, sample A, and at 0 V bias on the back gate followed a power law dependence on the density, $\mu \propto n^{\alpha}$, with $\alpha \sim 1.4$, at low density and a saturation of the mobility at high density. The decrease in mobility at all densities for the thinner quantum wells relative to the thicker well was interpreted to be due to an increase in interface roughness scattering, and provides evidence for interface roughness scattering being the dominant scattering mechanism in the thinner wells over the studied carrier density range.

The value of the maximum mobility at high carrier density was a sensitive function of the back gate voltage, as depicted by the comparison of the density dependence of the mobility for sample A at a fixed back gate voltage of 0 V and 0.8 V. At high density, under accumulation by application of a positive front gate voltage, application of a positive back gate voltage corresponded to a decrease in the electric field and a decrease in the potential asymmetry over the quantum well. The corresponding increase in mobility suggests a dominant scattering mechanism that is a function of the magnitude of the electric field and the position of the electron wave function over the quantum well; therefore, both interface roughness scattering and alloy scattering are suspected of being dominant scattering mechanisms at high electron densities, the latter being due to a non-uniform alloy distribution resulting from asymmetric element intermixing during MBE growth. The mobility was less sensitive to the potential asymmetry over the quantum well at low density, suggesting the contribution to carrier scattering by interface roughness and alloy disorder was lower at low carrier density.

The calculated mobility in Fig. 6(a) included contributions to scattering from rough interfaces under a self-consistent electric field, alloy disorder in the quantum well, homogeneous background impurities, two-dimensional remote impurities located at the dielectric/III-V interface, three-dimensional remote impurities located in the barriers, and charged dislocations. The functional dependence of mobility on density for sample C was found to be singularly dependent on interface roughness scattering, with fluctuation height $\Delta = 2.7$ Å and in-plane correlation length $\Lambda = 9$ nm, similar to previously reported values[25]. Having assumed the same interface roughness parameters apply to sample A and B, it was found that including alloy scattering with alloy composition $InAs_{0.987}Sb_{0.013}$ resulted in good agreement between the calculated and measured mobility for physically reasonable values of the electric field over the quantum wells.

Similar agreement between measurement and calculation could be achieved with adjustment to the interface roughness parameters by approximately +/- 10% of the aforementioned values while the alloy composition was adjusted by +/- 0.003. Neglecting either scattering mechanism did not yield agreement between calculations and measurements for all quantum well widths. The dependence of mobility at a carrier density of approximately 1.33 ($10^{12}$ cm$^{-2}$) on quantum well width is shown in Fig. 6(b) for sample A at back gate voltages ranging from 0 V to 1.0 V in increments of 0.2 V, and for samples B and C and a back gate voltage of 0 V. Calculated values of mobility are depicted by the curves for several magnitudes of the average electric field over the quantum well with the dashed curve indicating the case of zero electric field.

The remaining parameters used for scattering calculations were a homogeneous background impurity concentration, $N_{BI} = 1$ ($10^{14}$ cm$^{-3}$), remote impurities, $N_{2D,R} = 5$ ($10^{12}$ cm$^{-2}$) located at the dielectric/III-V interface, remote impurities, $N_{3D,R} = 1$ ($10^{17}$ cm$^{-3}$) distributed through the top and bottom barriers, and a threading dislocation density, $N_{disl}$ 1 ($10^{6}$ cm$^{-2}$). The chosen values of $N_{BI}$, $N_{2D,R}$, and $N_{3D,R}$ were within the range of experimental estimates of these values. Estimates of background impurities in the channel and barrier materials were based on Hall measurements of bulk layers and secondary ion mass spectrometry of impurities in heterostructures. An estimate of the density of scattering centers at the dielectric/III-V interface was determined by the gating efficiency and is in agreement with reported values for devices using a similar gate metallization technique[26]. The calculated functional dependence of the scattering rate on density by remote impurities, versus distributed remote impurities versus background impurities were similar, and therefore, the level of deviation between calculated and measured mobility similar to that generated for sample A could be obtained by simultaneously varying $N_{BI}$, $N_{2D,R}$, and $N_{3D,R}$. The agreement between calculated and measured mobility was generally better at low density for lower $N_{BI}$ and higher $N_{2D,R}$ and $N_{3D,R}$.

The density dependence of the mobility in samples D and E was qualitatively similar to that observed in sample A; however, the magnitude of the mobility at higher carrier density was slightly higher in samples D and E compared to sample A. The carrier density in sample D was increased through occupation of the second sub-band, and a corresponding drop in mobility was observed at carrier densities greater than ~1.7 ($10^{12}$ cm$^{-2}$). Fig. 7 depicts the density dependence of the mobility for samples D and E, the calculated mobility due to individual scattering mechanisms (solid lines), and the total calculated mobility (dashed line). The calculated mobility overlaying the data for samples D and E assumes a remote impurity concentration in the barriers, $N_{3D,R} = 8$ ($10^{16}$ cm$^{-3}$), an alloy composition of $InAs_{0.988}Sb_{0.012,}$ and interface roughness parameters, $\Delta = 2.7$ Å and $\Lambda = 13.0$ nm. The remaining parameters were identical to that used for sample A. The similarity in the mobility between sample A and samples D and E suggest shutter sequence I and shutter sequence II yield interfaces of similar quality, and the highly implemented and studied shutter sequence I may not be optimal. Further exploration of growth conditions for the heterointerfaces is merited and will likely be the source of higher mobility at higher carrier density in InAs quantum wells with lattice matched barriers.

At lower carrier density, samples A, D and E observe similar power law dependencies of the mobility on density, $\mu \propto n^{\alpha}$, with $\alpha \sim 1.3 - 1.6$. Such a power law dependence over the studied range of carrier density is consistent with mobility limited by coulomb scattering from impurities remote to the well[13]. As observed in sample A, in samples D and E increasing the coulomb scattering rate from impurities local to the well by significantly increasing the homogenous background impurity concentration relative to the value used for the calculations, $N_{BI} = 1 (10^{14} \text{ cm}^{-3})$, led to greater discrepancy between calculated mobility and measured values at low carrier density. However, the expected dependence of mobility on density is sensitive to screening of the scattering potential by the 2DEG[13], which itself is sensitive to the relative concentrations and locations of impurities.

Scattering from charged dislocations contributed little to the overall scattering rate. Accounting for scattering by remote impurities located in the barriers, $N_{3D,R}$, at concentrations of order $10^{17} \text{ cm}^{-3}$ led to reasonable agreement between calculated mobility and measured mobility at low carrier densities. A concentration of remote impurities in the barriers of the order $10^{17} \text{ cm}^{-3}$ is not unrealistic if both donor states and acceptor states are both present and in a ratio that would bring the total donor contribution to the quantum well in agreement with the measured sheet carrier densities. Donor levels may arise from $As_{Al}$ anti-site defects formed by un-intentional As incorporation in the barriers during MBE growth[16,28], and numerous intrinsic and extrinsic defects may form in the barriers. Electrically active intrinsic defects with the lowest reported formation energies[29] include interstitial aluminum, $Al_{i,Al}^{+1}$, antimony anti-sites, $Sb_{Al}^{+1}$, and aluminum vacancies $V_{Al}^{-3}$. Extrinsic defects from the common impurities carbon and oxygen are predicted[30] to form primarily acceptor levels from the substitutional defect $C_{Sb}$ and the interstitial defect $O_{i,tet,Al}$. Extrinsic and intrinsic defects of both donor and acceptor character are predicted to form over a wide range of chemical potential at temperatures close to the MBE growth temperature.

A final coulombic scattering mechanism is considered: that arising from a two-dimensional layer of charged defects contained at the interface between the barriers and quantum well. Such a model is physically meaningful if the $As_{Al}$ anti-site defects are formed predominantly near the interfaces. Not considering scattering from other mechanisms, a charge density of $10^{11}$ $cm^{-2}$ shared between the top and bottom interfaces would result in a calculated mobility in agreement with measured values at low carrier density.

The complexity of the system makes it difficult to pinpoint the origin(s) of coulombic scattering dominant at low density. However, the noted sources of charged defects suggest improvements to the quality of the AlSb barriers by optimizing growth conditions for individual epilayers and developing techniques to suppress unintentional group V intermixing and alloying may yield higher mobility at lower carrier densities. This work is forthcoming.

The growth and the density dependence of the low temperature mobility of multiple two-dimensional electron systems confined to un-intentionally doped, low extended defect density InAs quantum wells with $Al_{1-x}Ga_xSb$ barriers has been reported. A gate-tuned electron mobility exceeding 750,000 $cm^2/Vs$ was achieved at a sample temperature of 2 K. Analysis of the possible scattering mechanisms suggest that at high carrier density within the first occupied electronic sub-band, the mobility was limited by interface roughness and alloy scattering. At low carrier density, the functional dependence of the mobility on carrier density and gate voltage provided evidence of coulombic scattering from charged defects.

*Acknowledgements*

The authors thank David D. Awschalom for assistance with measurements. This work was supported by Microsoft Research Station Q. This work made use of the central facilities of the UCSB MRL, which is supported by the MRSEC program of the National Science Foundation under Award No. DMR-1121053. This work also made use of the UCSB Nanofabrication Facility, a part of the NSF funded NNIN network, and of the California NanoSystems Institute.

### Appendix A: Model and theory for calculating electron mobility

The system considered was a two-dimensional electron gas confined to an InAs quantum well with $Al_xGa_{1-x}Sb$ barriers. The electrons are mobile in the xy plane and are confined in the z direction. The envelope wave function, $\varphi(z)$, over the InAs layer of thickness L was approximated by

$$\varphi(z) = \left(\frac{2}{L}\right)^{1/2} \sin\left(\frac{\pi z}{L}\right), 0 \leq z \leq L \tag{A1}$$

and zero for all other z. The subband structure was neglected; only the lowest occupied subband was treated. The electron mobility $\mu = e\tau/m^*$, where $e$ is the elementary charge, is a function of the momentum relaxation time, $\tau$, and the electron effective mass, $m^*$. An estimate of the effective mass accounting for band non-parabolicity[31] was used in the proceeding calculations.

Under the Born approximation the general form of the momentum relaxation time is given by[24]

$$\frac{1}{\tau_i} = \frac{1}{2\pi\varepsilon_f} \int_0^{2k_f} dq \, \frac{q^2}{\sqrt{4k_f^2 - q^2}} \frac{\langle |U_i(q)|^2 \rangle}{\epsilon_q^2}, \tag{A2}$$

where the integration is over wave number, $q$, and the subscript $i$ labels the scattering mechanism under consideration. Within the random-phase approximation the dielectric matrix, $\epsilon_q$, is given by

$$\epsilon_q = 1 + V(q)[1 - G(q)]X^0(q) \, , \tag{A3}$$

where $G(q)$ is the Hubbard form of the local-field correction and $X^0(q)$ is the polarizability of the 2DEG. The electron-electron interaction $V(q)$ is characterized by a form factor and Coulomb potential due to the finite confinement and is expressed as

$$V(q) = \frac{q_s}{q} F_c(q) \, , \tag{A4}$$

where $q_s = \frac{2\pi \mathrm{e}^2}{\epsilon_L}$ is the screening parameter with dielectric constant, $\epsilon_L$, and $F_c(q)$ is given by

$$F_c(q) = \int_{-\infty}^{+\infty} dz |\varphi(z)|^2 \int_{-\infty}^{+\infty} dz' |\varphi(z')|^2 \exp(-q|z - z'|) \, . \tag{A5}$$

$\langle |U_i(q)|^2 \rangle$ is the averaged random potential corresponding to specific forms of defects leading to elastic scattering.

Rough quantum well interfaces and the presence of an electric field results in an averaged random potential, $\langle |U_{IR}(q)|^2 \rangle$, of the form[22]

$$\langle |U_{IR}(q)|^2 \rangle = \pi \mathrm{F}^2 \Delta^2 \Lambda^2 \exp(-(q\Lambda)^2/4) \, . \tag{A6}$$

Here, F is a function that includes terms for the variation in the quantum well width and a shift in the ground state energy in the presence of an electric field, E, from a perturbative treatment:

$$\mathrm{F} = -\left( \frac{\hbar^2 \pi^2}{m^* L^3} + 96 \left( \frac{2}{3\pi} \right)^6 \frac{e^2 m^* L^3 E^2}{\hbar^2} \right) \, . \tag{A7}$$

The quantum well width fluctuation due to interface roughness is parameterized by the height $\Delta$ and the in-plane correlation length $\Lambda$.

Un-intentional group-V (group-III) intermixing between the well and barriers during MBE growth can lead to the formation of InAs$_x$Sb$_{1-x}$ (In$_x$Al$_{1-x}$As) within the quantum well. The case of alloy scattering within the quantum well due to group-V intermixing was considered. Perfectly random alloy disorder results in a short range fluctuating potential, and the averaged random potential, $\langle |U_{alloy}|^2 \rangle$, is expressed as[23]

$$\langle |U_{alloy}|^2 \rangle = x(1-x)\delta V^2 \Omega \int dz\, \varphi(z)^4 \ . \tag{A8}$$

Here, $\delta V$ is the spatial average of the fluctuating alloy potential, and $\Omega = \frac{\sqrt{3}\pi}{16}a^3$ is the volume of scattering potential in an alloy with lattice parameter, $a$. The spatial average of the fluctuating alloy potential, $\delta V = 0.8$ eV, was determined for the specific alloy composition from the heteropolar energy associated with the dielectric method of calculating the bandstructure.[32]

Remote ionized impurities confined to a two-dimensional plane lead to a random potential, $\langle |U_{R,2D}(q)|^2 \rangle$, of the form[20]

$$\langle |U_{R,2D}(q)|^2 \rangle = \left( \frac{2\pi e^2}{\epsilon_L} \frac{1}{q} \right)^2 n_i(z) F(q, z_i)^2 \ , \tag{A9}$$

where $n_i(z)$ is the impurity concentration. The form factor $F(q, z_i)$ accounts for the finite width of the quantum well with the distance $z_i$ between the impurity layer and the quantum well and is given by

$$F(q, z_i) = \int_{-\infty}^{+\infty} dz\, |\varphi(z)|^2 \exp(-q|z - z_i|) \ . \tag{A10}$$

The form of the average random potential, $\langle |U_{R,2D}(q)|^2 \rangle$, was extended for calculating the momentum relaxation time due to remote ionized impurities distributed in three dimensions by integration.

Homogeneous background impurities lead to a random potential[20]

$$\langle |U_B(q)|^2 \rangle = \left( \frac{2\pi e^2}{\epsilon_L} \frac{1}{q} \right)^2 N_B L F_B(q) \, ,$$

(A11)

where $N_B$ is the concentration of the three dimensional background ionized impurities in the InAs quantum well with the form factor

$$F_B(q) = \frac{1}{L} \int_{-\infty}^{+\infty} dz_i F(q, z_i)^2 \, .$$

(A12)

The random potential, $\langle |U_{disl}(q)|^2 \rangle$, due to charged dislocations treated as a line charge[21] with charge density $\rho_l$ along the dislocation line and areal density $N_{disl}$ is given by

$$\langle |U_{disl}(q)|^2 \rangle = N_{disl} \left( \frac{e}{\epsilon_L} \frac{\rho_l}{q} \right)^2 F_d(q),$$

(A13)

where the form factor is given by

$$F_d(q) = \int_{-\infty}^{+\infty} dz_i F(q, z_i)^2.$$

(A14)

The total momentum relaxation time was determined from the individual relaxation times associated with each scattering mechanism by application of Mathiessen's rule,

$$1/\tau = 1/\tau_{IR} + 1/\tau_{alloy} + 1/\tau_{R,2D} + 1/\tau_{R,3D} + 1/\tau_B + 1/\tau_{disl}.$$

(A15)

*Figures and tables*

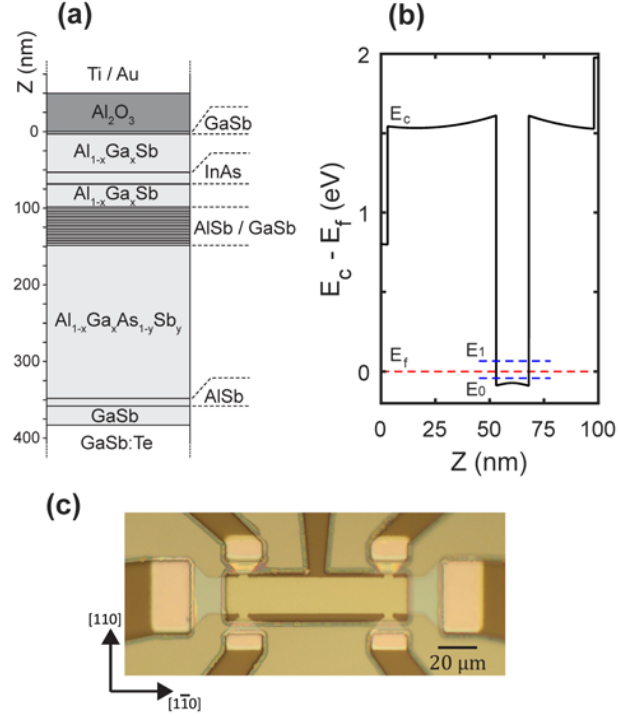

**Figure 1. (a)** A schematic of the InAs quantum well heterostructure with Al$_2$O$_3$ gate dielectric and Ti/Au front gate. **(b)** A self-consistent calculation of the conduction band profile and the first and second subband levels relative to the Fermi level for an Al$_{0.8}$Ga$_{0.2}$Sb/InAs/ Al$_{0.8}$Ga$_{0.2}$Sb quantum well. **(c)** An optical micrograph of one of the Hall bar devices used in this study prior to gate and ohmic metallization.

**Table 1.** Characteristics of the InAs quantum wells discussed in the text.

| Sample | InAs thickness (nm) | Barrier composition | Interface formation | †N$_{Hall}$ ($10^{11}$ cm$^{-2}$) | †N$_{SdH}$ ($10^{11}$ cm$^{-2}$) | †$\mu$ (cm$^2$/V·s) |
|--------|---------------------|---------------------|---------------------|-----------------------------------|----------------------------------|---------------------|
| A | 15 | AlSb | Procedure I | 3.81 | 3.82 | 258,000 |
| B | 10 | AlSb | Procedure I | 4.64 | 4.53 | 117,000 |
| C | 8 | AlSb | Procedure I | 3.59 | 3.44 | 20,000 |
| D | 15 | AlSb | Procedure II | 4.47 | 4.64 | 332,000 |
| E | 15 | Al$_{0.8}$Ga$_{0.2}$Sb | Procedure II | 4.43 | 4.69 | 355,000 |

† Measured at a sample temperature of 2 Kelvin and V$_f$ = V$_b$ = 0 V

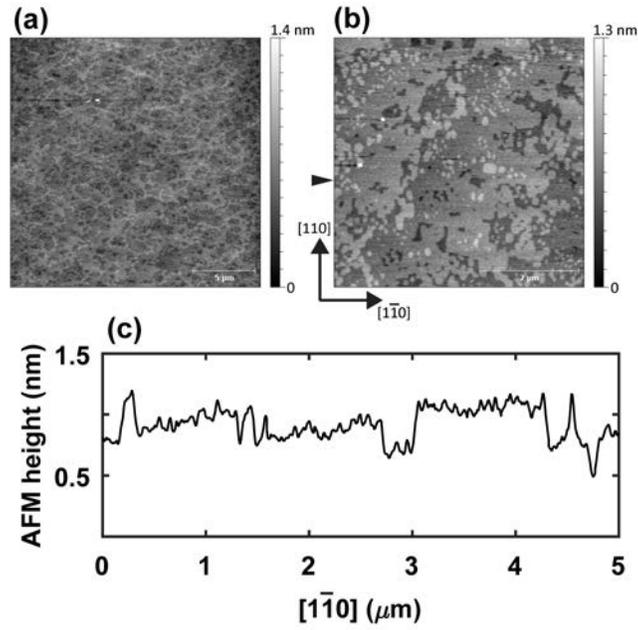

**Figure 2. (a)** A 20 x 20 µm$^2$ atomic force microscope image of sample E taken at the surface of the GaSb capping layer. **(b)** A 5 x 5 µm$^2$ subset of the image in (a). **(c)** A line scan along the [1-10] crystallographic direction with origin at the arrow in (b).

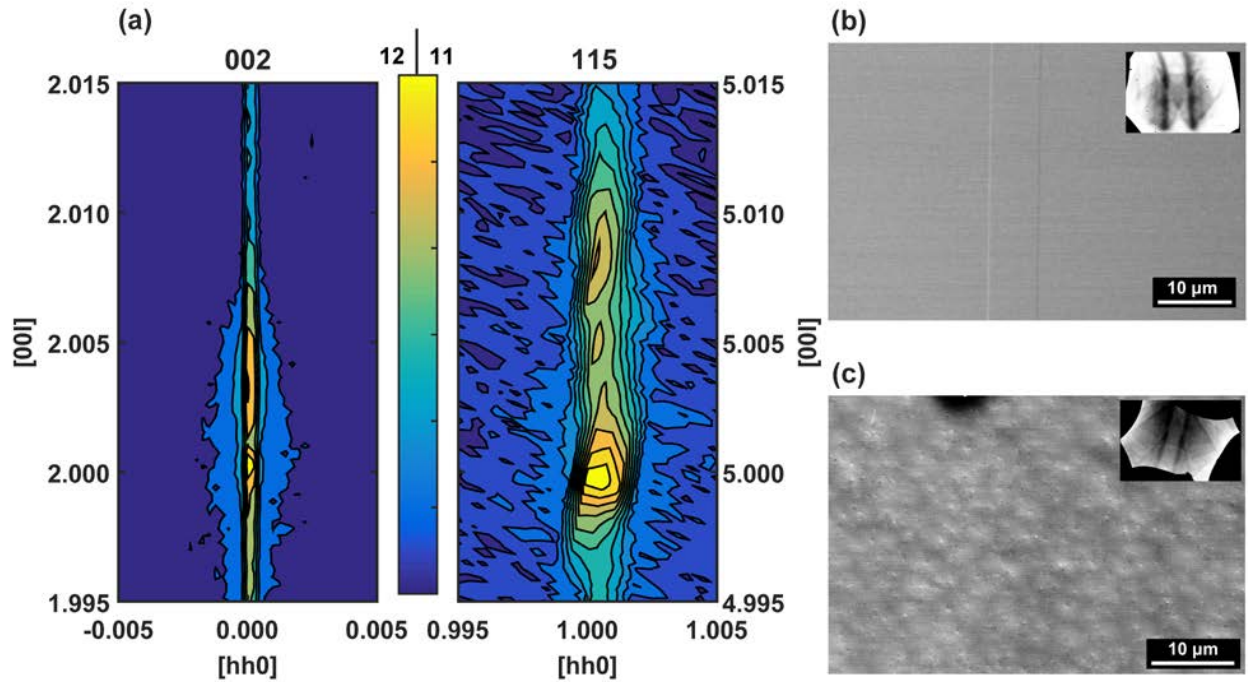

**Figure 3. (a)** X-ray diffraction reciprocal space maps of sample D measured close to the 002 and close to the asymmetric 115 Bragg reflections of the GaSb substrate. The color scale linearly spans the logarithm of the intensity. **(b)** Electron channeling contrast image of sample D. The inset shows the channeling pattern. **(c)** Electron channeling contrast image of a reference GaSb buffer grown on a lattice mismatched GaAs (001) substrate. The inset shows the channeling pattern.

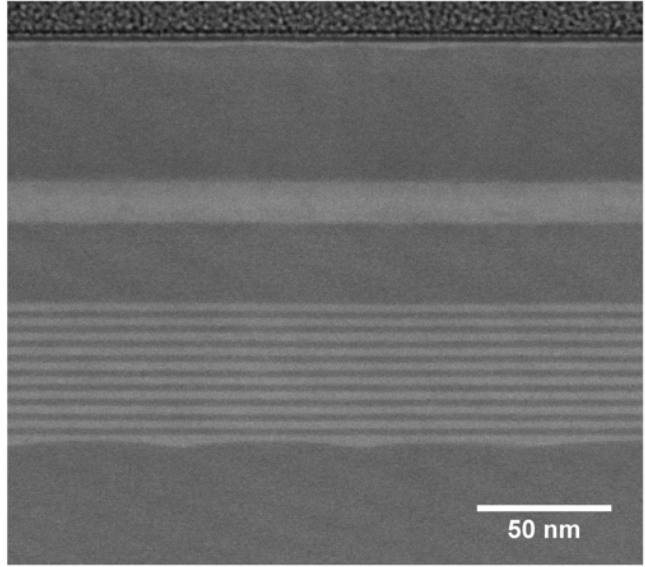

**Figure 4**. High angle annular dark field (HAADF) scanning transmission electron microscope (STEM) image of sample D along the [110] zone axis and in the vicinity of the InAs quantum well. The 80 nm thick specimen was prepared by milling a section of wafer using a focused gallium ion beam.

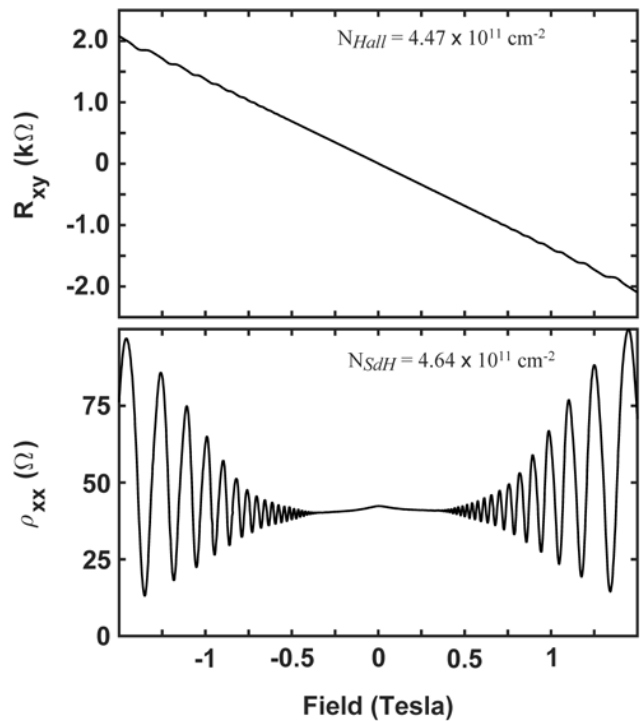

**Figure 5.** The Hall resistance (upper plot) and magnetoresistivity (lower plot) of sample D at $V_f = V_b = 0$ V and at a sample temperature of 2 Kelvin.

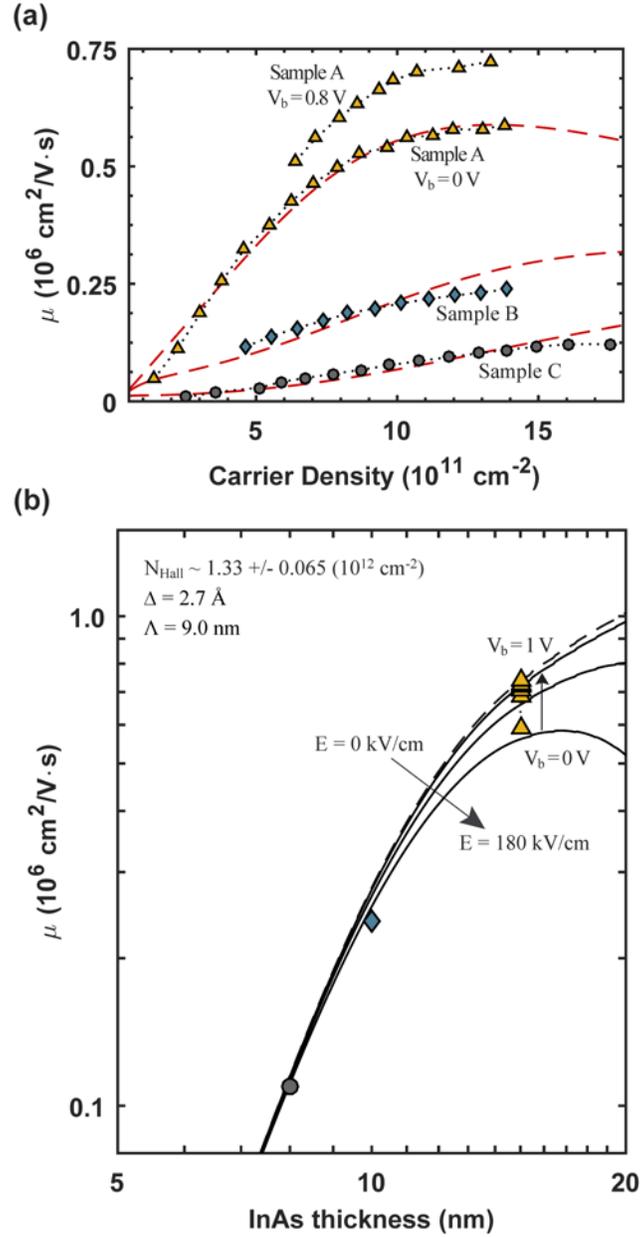

**Figure 6. (a)** The dependence of mobility on carrier density for sample A at constant back gate voltages $V_b = 0$ V and 0.8 V, and for sample B and C at constant back gate voltage $V_b = 0$ V. The carrier density was changed by varying the front gate, $V_f$. The red dashed curves depict the calculated total mobility. **(b)** The dependence of mobility on the InAs thickness. The calculated mobility for zero average electric field over the quantum well is shown as the dashed line. The mobility for electric fields of 60, 120 and 180 kV/cm are shown as solid lines. The carrier density

determined by Hall measurements and the interface roughness parameters used in the calculations are shown in the inset. The remaining parameters were and alloy composition of InAs$_{0.987}$Sb$_{0.013}$, a homogeneous background impurity concentration, N$_{BI}$ = 1 ($10^{14}$ cm$^{-3}$), remote impurities, N$_{2D,R}$ = 5 ($10^{12}$ cm$^{-2}$) located at the dielectric/III-V interface, remote impurities, N$_{3D,R}$ = 1 ($10^{17}$ cm$^{-3}$) distributed through the top and bottom barriers, and a threading dislocation density, N$_{disl}$ 1 ($10^{6}$ cm$^{-2}$).

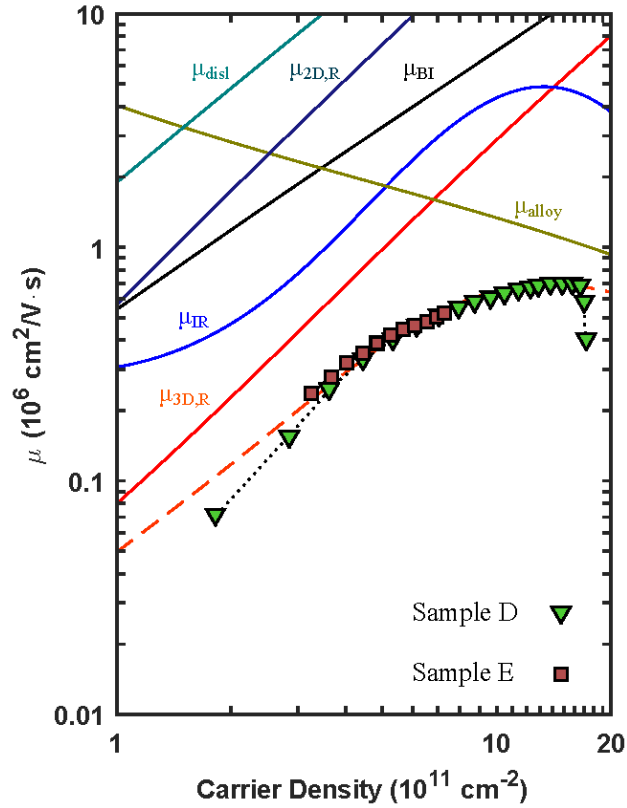

**Figure 7.** The dependence of mobility on carrier density for sample D and E at constant back gate voltage V$_b$ = 0 V. The carrier density was changed by varying the front gate, V$_f$. Mobility limited by individual scattering mechanisms are depicted by solid lines with parameters for interface roughness Δ = 2.7 Å and Λ = 13.0 nm, a self-consistent electric field, an alloy composition of InAs$_{0.988}$Sb$_{0.012}$, a homogeneous background impurity concentration, N$_{BI}$ = 1 ($10^{14}$ cm$^{-3}$), remote impurities, N$_{2D,R}$ = 5 ($10^{12}$ cm$^{-2}$) located at the dielectric/III-V interface, remote impurities, N$_{3D,R}$ = 8 ($10^{16}$ cm$^{-3}$), distributed through the top and bottom barriers, and a threading dislocation density, N$_{disl}$ 1 ($10^{6}$ cm$^{-2}$). The dashed red line indicates the calculated total mobility.